\documentclass[conference,onecolumn]{IEEEtran}

\usepackage[utf8x]{inputenc}
\usepackage[table,xcdraw]{xcolor}
\usepackage[pdfa]{hyperref}
\usepackage{balance}
\usepackage{cite}
\usepackage{csquotes}
\usepackage[english]{babel}
\usepackage{amsmath,amssymb,amsfonts}
\usepackage{algorithmic}
\usepackage{graphicx}
\usepackage{textcomp}
\usepackage{xcolor}
\usepackage{jabbrv}
\def\BibTeX{{\rm B\kern-.05em{\sc i\kern-.025em b}\kern-.08em
T\kern-.1667em\lower.7ex\hbox{E}\kern-.125emX}}
\usepackage[inline,shortlabels]{enumitem}

\usepackage{xcolor}
\definecolor{added}{rgb}{0.65, 0.25, 0.39}
\newcommand{\added}[1]{{\color{added}#1}}
\usepackage{multicol}

\begin{document}

\title{Audio-to-Score Alignment Using \\ Deep Automatic Music Transcription}

\author{
	\IEEEauthorblockN{Federico Simonetta, Stavros Ntalampiras, Federico
		Avanzini}
	\IEEEauthorblockA{
		LIM --- Music Informatics Laboratory\\
		Department of Computer Science\\
		University of Milan\\
		Email: \{name.surname\}@unimi.it    }
}

\maketitle

\added{
	\Large \textbf{
		This PDF is an updated version of the paper published at the
		IEEE MMSP 2021. It contains some erratum highlighted in red, especially in
		sections~\ref{sec:experimental_setup},~\ref{sec:results},~and~\ref{sec:conclusions}.
	}
}

\begin{multicols}{2}

	\begin{abstract}
		Audio-to-score alignment (A2SA) is a multimodal task consisting in the
		alignment of audio signals to music scores. Recent literature confirms the
		benefits of Automatic Music Transcription (AMT) for A2SA at the frame-level.
		In this work, we aim to elaborate on the exploitation of AMT Deep Learning
		(DL) models for achieving alignment at the note-level. We propose a method which
		benefits from HMM-based score-to-score alignment and AMT, showing a
		remarkable advancement beyond the state-of-the-art. We design a
		systematic procedure to take advantage of large datasets which do not offer
		an aligned score. Finally, we perform a thorough comparison and extensive
		tests on multiple datasets.
	\end{abstract}


	\section{Introduction}

	Audio-to-score alignment (A2SA) is a Music Information Retrieval
	(MIR)
	task which aims at finding correspondences between
	time instants in a music recording and time instants in the associated music
	score. Such a technology facilitates various tasks, ranging from cultural
	heritage applications attempting to ease the fruition of music, to preprocessing
	stage for various multimodal MIR tasks~\cite{simonetta201901multimodal}.

	A major difference in A2SA methods is set between online and offline alignment.
	Online methods, often named ``score-followers'', try to predict the time instant
	in which a new note is played and track the change without future information about the performance.
	Offline methods, instead, try to match time instants by exploiting the knowledge
	of the full performance. In this work, we will concentrate on offline A2SA.

	Similarly to other alignment problems, offline A2SA can be addressed using dynamic
	programming approaches and, as such, most of the literature focuses on Dynamic Time
	Warping (DTW) based methods~\cite{muller2007dynamic}. DTW is an algorithm which
	is able to find the minimum cost path in a fully connected graph where nodes are
	the elements of two sequences and branches are weighted according to a given
	distance function. Even though DTW is effective and versatile, it has a strong
	requirement: the two input sequences must be sorted with the same element order.
	Formally, given any two pairs of corresponding elements $(a', b')$ and $(a'',
		b'')$, then $a' \geq a'' \implies b' \geq b''$.
	This requisite is met by music representations at sample or
	frame-level, but it hinders the alignment of polyphonic music at the note-level
	because the sequence of note onsets and offsets in a performance is not always
	the same as in the score. As consequence, most DTW methods use a sequence of frames
	as input. Moreover, since DTW is based on a distance function, such methods
	focus on discovering the optimal function and feature space.

	There is a limited number of works that have faced the problem of note-level
	alignment, mainly with the objective of music performance analysis. Indeed, it
	is known that subtle asynchronies are generated during a human performance:
	notes in the same chord are written in musical scores as events having the same
	onset -- and possibly the same offset --, but music players always introduce
	asynchronies of less than 0.05 seconds among the timings of such
	notes~\cite{devaney201407estimating}.
	Other discrepancies between score and performance note order are related to
	the phrasing and articulation practices; for instance, the \textit{legato}
	articulation consists in a slight overlap between two successive notes, even if
	in the musical score they are notated with no overlap. These almost
	imperceptible timing effects are considered to be responsible of the incredibly
	various expressiveness of music performances and are consequently of crucial
	importance in music performance analysis studies~\cite{lerch2019music}. Methods
	used for note-level alignment so far include HMM, DTW, NMF and blob recognition
	in spectrograms~\cite{carabiasorti201510audio, miron2014audiotoscore,
		wang201504compensating, devaney201407estimating, wang2018scorealigned}

	The rise of Artificial Neural Networks in their Deep Learning (DL)
	paradigm has led several researchers to exploit DL models for feature learning
	tailored to DTW.\@ Two methods are particularly noteworthy:
	one~\cite{agrawal2021learning} employs Siamese Networks for learning
	features  that can be used for some distance function in DTW;\@ the
	second method relies on the improvements made in the field of AMT for
	converting the sequences to a common space --- the space of the symbolic
	notation~\cite{kwon2017audiotoscore}.

	In this work, we elaborate on the exploitation of AMT DL-based models for
	achieving note-level alignment. We propose a method which benefits
	from HMM-based score-to-score alignment and AMT, showing a remarkable
	advancement against the state-of-the-art. We design a systematic procedure to
	take advantage of large datasets, where aligned scores are unavailable.
	Last but not least, we perform a thorough comparison and extensive tests on
	multiple datasets.
	For reproducibility purposes, the implementation of the proposed method along with the present
	experiments is available
	online.\footnote{\url{https://github.com/LIMUNIMI/MMSP2021-Audio2ScoreAlignment}}

	\begin{figure*}
		\center
		\includegraphics[height=0.75\textheight]{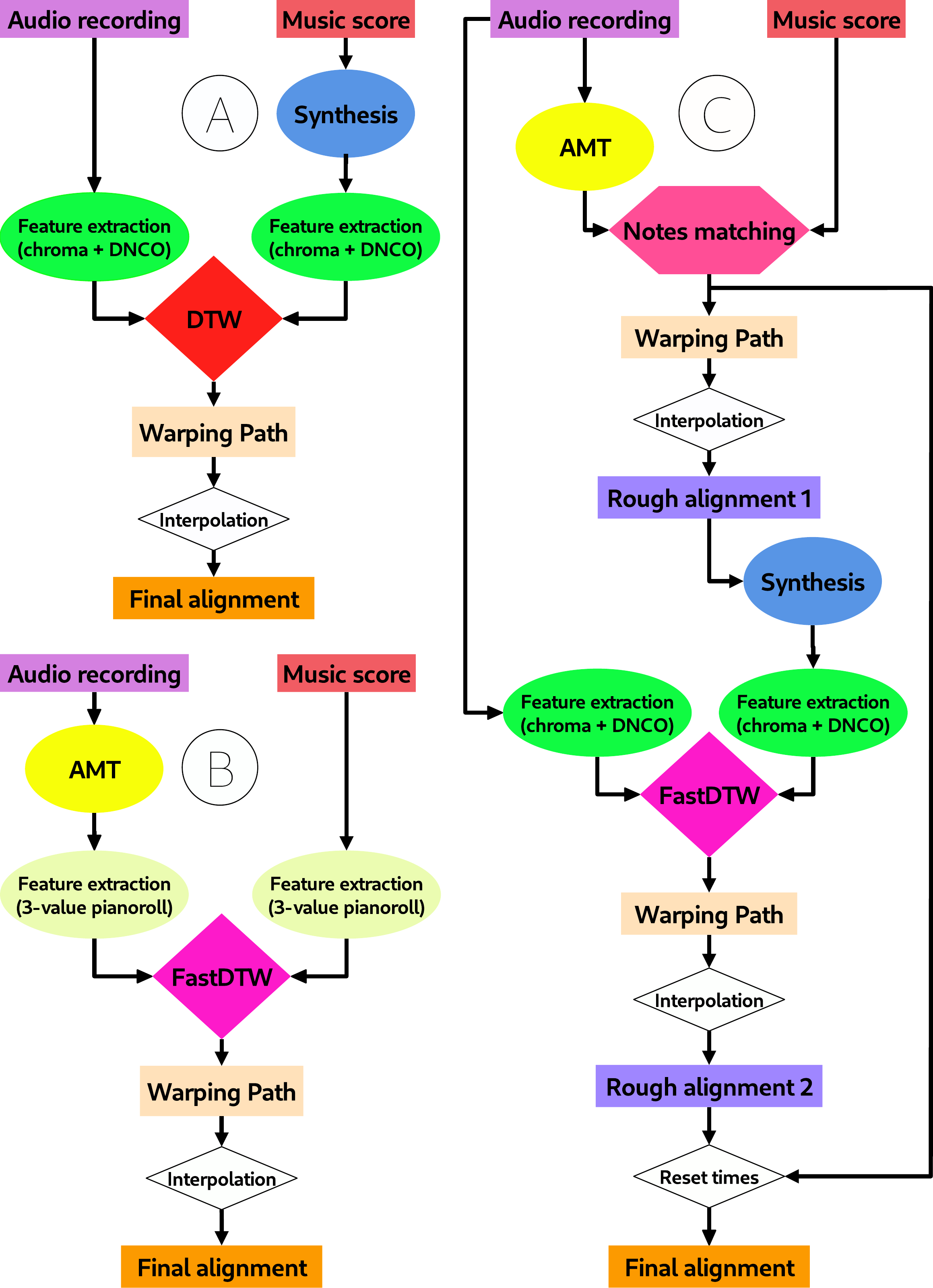}
		\caption{Flow chart of the three methods used here: \textcircled{\scalebox{0.7}{A}} SEBA
			method; \textcircled{\scalebox{0.7}{B}} TAFE method; \textcircled{\scalebox{0.7}{C}} EIFE method }
		\label{fig:scheme}
	\end{figure*}

	\section{Baseline method}
	DTW requires as input a distance matrix representing every possible matching
	between sorted sequence elements. If $N$ and $M$ are
	the number of elements in the input sequences, the distance matrix will have
	size $N \times M$. DTW finds the shortest path from element $(1, 1)$ to $(N, M)$ according to so-called \textit{local} and \textit{global}
	constraints. \textit{Local} constraints list all possible moves among which
	the algorithm can chose during the computation of the path, while
	\textit{global} constraints limit the computational complexity of the procedure
	(which in the no-constrained form is dominated by $O(M \times N)$ in both time and
	memory). As a consequence, DTW is highly expensive for long sequences: for
	instance, to align sample-to-sample two audio recordings lasting 10 minutes with
	sample rate 22050 Hz, DTW needs a distance matrix with $1.75 \times 10^{14}$
	elements, meaning $318$ Terabytes using 16-bit floating point numbers.

	Apart from the \textit{global} constraints, various approximated alternatives
	for common local constraints have been proposed to relax the high complexity in
	time and memory, with \textit{FastDTW}~\cite{salvador2007toward} being one of
	the most widely adopted solutions.

	Interestingly, one of the most widespread methods for A2SA consists in
	converting all the data to the audio level (usually by synthesizing the music
	score) and subsequently extracting some audio-related features (see
	Figure~\ref{fig:scheme}\textcircled{\scalebox{0.7}{A}}). Notably, one
	method~\cite{ewert200904resolution} uses the sum of two distance matrices
	computed with two different combinations of audio features and distance
	functions; the main objective is to consider both percussive and harmonic
	features of musical instrument acoustics. In this paper, we will refer to such
	method with the name SEBA\footnote{Here and in the following sections, SEBA,
		EITA, TAFE and, EIFE refer to the first syllables of the researcher first
		name who worked at the corresponding method --- i.e.
		SEBAstian~\cite{ewert200904resolution}, EITA\cite{nakamura2017performance},
		TAegyun~\cite{kwon2017audiotoscore} and FEderico (this paper first author)}.

	\section{The proposed alignment methods}

	\subsection{AMT-based frame-level alignment}

	AMT consists in the analysis of music audio recordings to discover semantically
	meaningful events, such as notes, instruments and chords. Two main methodologies
	for AMT exist, i.e. Non-negative Matrix Factorization (NMF) and Deep Learning
	(DL) (for a thorough review see~\cite{benetos201901automatic}). During the last
	4 years, DL has tremendously advanced the state-of-art of AMT, especially for
	piano music recordings~\cite{hawthorne2018onsets,kong2020highresolution}. Due to
	the high variability of timbres, instrumental acoustics, playing practice, and
	difficulties in collecting data, multi-instrument AMT remains a hard challenge.

	To our knowledge, the state-of-art of A2SA for piano
	music~\cite{kwon2017audiotoscore} is based on \begin{enumerate*}[(1)] \item AMT
		of recorded audio,
		and, \item alignment of piano-roll representations of music \end{enumerate*}.
	This approach can be seen as the opposite of classical DTW methods since
	it converts data to the symbolic domain instead of the audio domain.

	Piano-rolls are 2D boolean matrices with $K$ rows and $N$ columns in which the
	entry $(k, n)$ is 1 if pitch $k$ at time $n$ is playing, and 0 otherwise.
	Usually, $K$ is set to $128$ so that it is directly related to the MIDI
	specifications.

	In~\cite{kwon2017audiotoscore}, an AMT system is used to infer a MIDI
	performance; from there, a piano-roll is constructed. Piano-rolls coming from the
	transcribed audio and from the score can then be aligned using FastDTW to create
	a mapping between columns (frames) in the score domain and columns in the audio
	domain.  The mapping, so-called ``warping path'', can be used to recompute the
	correct duration of the notes found in the score without relying on the AMT
	output, which is prone to errors in pitch identification~\cite{benetos201901automatic}.

	Following the same line of thought, we used the new state-of-art piano AMT
	model~\cite{kong2020highresolution} for the alignment, which by itself has a
	greater precision than the method previously used. The second improvement we
	made is the use of 3-valued piano-rolls, that is, we introduced a new value to
	represent the onset of a note bringing two advantages: \begin{enumerate*}[(1)]
		\item in
		the boolean piano-roll, repeated notes are not distinguishable if the
		onset is
		immediately after the offset of the previous note, and, \item the
		introduction
		of a new value works as ``anchor'' for the DTW algorithm, which tries to
		find
		correspondences between the alternations of three values instead of only
		two
	\end{enumerate*}. We also attempted to use the same approach for
	multi-instrument A2SA by using a state-of-art multi-instrument AMT
	model~\cite{wu2020multiinstrument}.

	Here, we will refer to the specific method with the name \textit{TAFE}.
	A schematic
	representation of the method is depicted in Figure~\ref{fig:scheme}\textcircled{\scalebox{0.7}{B}}.

	\subsection{AMT-based note-level alignment}

	The merit of the above method was to highlight that DL models for piano AMT
	are tremendously effective in identifying onsets; however, the accuracy with
	pitch identification is low, due to issues such as false octaves and fifths.
	Moreover, as we will show in the results, offset time identification is still an
	unresolved obstacle.

	On the other side, the TAFE method relies on DTW algorithm for aligning
	score and audio at the frame-level. This not only is a high computationally
	demanding task, but also suffers from the DTW requirements; in other words, it cannot align transcribed notes to score
	notes because of the discrepancies between score note order and audio note order. As such, it fails in two important tasks:

	\begin{itemize}
		\item it cannot handle correctly the subtle asynchronies that a human
		      performer introduces among onsets of notes in the same chord and that
		      are fundamental for performance analysis~\cite{lerch2019music};
		\item it cannot correctly align scores that differ from the recorded
		      performance or from the transcribed one --- e.g.\ has some missing/extra
		      note(s).
	\end{itemize}

	In the music alignment domain, a few studies have faced the problem of
	aligning scores and music performances which refer to the same music
	piece but differ in terms of presence/absence of a few notes --- e.g.\ wrong
	performances, different score editions, etc. Such methods usually try to
	classify if a certain note in a score/performance is a \textit{missing} note or an
	\textit{extra} note compared to another score/performance. Commonly used approaches
	are DTW and HMM, with the latter being so far the best-performing
	approach~\cite{nakamura2017performance}.

	To overcome the TAFE issues, we propose to use what we here call \textit{EITA}
	method\cite{nakamura2017performance}. EITA uses HMM to create a mapping
	between two sequences of notes in order to identify missing/extra notes and
	notes that have different pitches --- e.g.\ wrong pitch inferred by the AMT.\@

	However, after having identified extra notes in the performance/transcription,
	missing notes remain in the score. Here, we assume that such notes were
	actually played but not identified by the AMT. In this perspective, we build a
	warping path from the mapping between notes matched by EITA in score and AMT
	output; then, we use the warping path to linearly interpolate the onsets and
	offsets times of the non-matched notes --- i.e.\ notes in the score that are not
	present in the transcription according to EITA. To align such notes with even
	higher precision, we apply the standard SEBA method based on the synthesis of the
	score.
	Moreover, to reduce the computational cost, we use FastDTW instead of the
	classic version. Since we expect that the AMT output contains precise onsets,
	after SEBA processing, we set the EITA matched notes onsets using AMT output and
	keep SEBA alignment only for non-matched notes. The flow-chart of this
	method is shown in Figure~\ref{fig:scheme}\textcircled{\scalebox{0.7}{C}}. In the next, this method is
	referred to as \textit{EIFE}.

	\section{The Employed Datasets}
	\label{sec:datasets}

	Unfortunately, there is not a great variety of datasets providing exact matches
	between score and midi performances. Thus, we used a systematic approach to
	generate misaligned sequences of notes as similar as possible to a musical
	score. The drawback of our method is that the resulting evaluation will not
	produce reliable values for real-world applications. However, it ensures that
	data does not contain manual annotation errors regarding matching notes.
	Moreover, here we are interested in the comparison of the considered approaches
	and leave the perceptual assessment of a performance on real-world score for
	future work.

	In our previous work \cite{simonetta2020automatic}, we proposed a simple way for
	statistically modeling misalignments between scores and performances, and used
	such models to recreate similar misalignments for datasets not including scores,
	collecting them in framework called ``ASMD''. Here, we
	improve upon it by using meaningful statistics and inference. This
	section will also work as scientific reference for the new version of ASMD.

	The first improvement we made is the addition of the
	ASAP~\cite{foscarin2020dataset} dataset to enlarge the number of considered
	statistics. Second, we used the EITA method to select matching notes against
	which we compute statistics as well. Third, instead of modeling the misalignment
	of onsets and offsets, we have now recorded statistics about the onsets and the
	duration ratio between score and performance. Fourth,
	statistics are computed with models trained on the ``stretched'' scores, so that
	the training data consists of scores at the same average BPM as the performance;
	as such,
	the misaligned data consists of times at that average BPM.


	More precisely, we create statistical models as follows:

	\begin{enumerate}
		\item we compute standardized onset misalignment and duration ratio for each
		      note by subtracting the mean value for that piece and dividing by the standard deviation;
		\item we collect two histograms, one for the standardized onset misalignments ($X_{ons}$) and one for the standardized duration ratios ($X_{dur}$);
		\item we collect each piece-wise mean and standard deviation in four
		      histograms: two for the onset misalignment means and standard deviations
		      ($Y_{ons}^m$, $Y_{ons}^{std}$), and two for duration ratio means and standard deviations ($Y_{dur}^m$, $Y_{dur}^{std}$)
	\end{enumerate}

	To infer a new misaligned onset or duration, we choose a standardized value for
	each note from histograms $X_{ons}$ and $X_{dur}$, and a mean and a
	standard deviation for each piece, using the corresponding histograms $Y_{ons}^m$, $Y_{ons}^{std}$, $Y_{dur}^m$, $Y_{dur}^{std}$; with
	these data, we compute a non-standardized onset misalignment and duration ratio for each note. These two latter values can be used in reference to the ground-truth performance to compute the misaligned timing values.

	\begin{table}
		\center
		\caption{L1 macro-average error for artificial misalignments.}
		\begin{tabular}{l|c|c|}
			\cline{2-3}
			                                          & \textbf{Ons} & \textbf{Offs}
			\\
			\hline
			\multicolumn{1}{|l|}{\textbf{GMM-HMM}}    & 18.6 ± 49.7  & 20.7 ± 50.6   \\
			\hline
			\multicolumn{1}{|l|}{\textbf{Histograms}} & 7.43 ± 15.5  & 8.95 ± 15.5   \\
			\hline
		\end{tabular}
		\label{tab:hmm}
	\end{table}

	We actually tested two methods for choosing the standardized value: an HMM with
	Gaussian mixture emissions (GMM-HMM) and the above-described histogram-based
	sampling. We hand-tuned the HMMs finding an optimum in 20 states and 30 mixture components for onsets and 2 states
	and 3 components for duration models. We then compared GMM-HMM and histogram
	models on the notes matched by the EITA method. During this evaluation, we used
	the scores provided by ASMD for a total of 875 scores, namely ``Vienna
	Corpus''~\cite{goebl1999vienna}, ``Traditional
	Flute''~\cite{brum2018traditional},
	``MusicNet''~\cite{thickstun2018invariances},
	``Bach10''~\cite{duan2011soundprism} and ``asap''~\cite{foscarin2020dataset}
	group from ``Maestro'~\cite{hawthorne2019enabling} dataset.
	We divided the data into train and test sets with 70-30 proportion, resulting in
	641 pieces for training and 234 for testing. As evaluation measure, we used the
	L1 macro-average error between
	matching note onsets and offsets in music score and performance. However, due to
	EITA's high computational cost, we removed the scores for which EITA terminates after
	20 seconds. This resulted in a total of $347$ songs for training and $143$ songs
	for testing --- $\sim$54\% and $\sim$61\% of the corresponding splits.
	Table \ref{tab:hmm} shows the results.

	Misaligned data are finally created, using the histogram-based method for every
	dataset provided by ASMD by collecting the histograms corresponding to all 875
	scores --- 481 considering songs where EITA method took less than 20 sec.
	Thus, we set up a corpus of 1787 music recordings with misaligned and aligned
	MIDI data.

	Artificially misaligned data is more similar to a different performance than to
	a symbolic score; however, for most of MIR applications, such misaligned data is
	enough  to cover both training and evaluation needs. To achieve an even more
	accurate evaluation, in this work we also applied a single-linkage clustering to
	the onsets of each misaligned score. We stopped the agglomerative procedure when
	a certain minimum distance $t$ among clusters was reached. We have randomly
	chosen such threshold in $[0.03, 0.07]$ seconds, representing broad interval
	around 0.05 seconds that is assumed as upper-bound of usual chord
	asynchronies~\cite{devaney201407estimating}. Subsequently, we set the onsets of
	the notes in each cluster equal to the average onset time of that cluster so
	that the final misaligned note sequence contains chords made by notes having the
	same onset. This is a crucial difference between scores and performance data, in
	which chords are played with light asynchronies between same-onset notes.

	In the updated version of ASMD, we provide randomly generated missing and extra
	notes as well. To this end, we chose the number $n$ of notes to be tagged
	as ``missing'' or ``extra'' as a random variable with uniform distribution in
	$(0.1 \times L, 0.5 \times L)$, where $L$ is the number of notes in the music
	piece. Then, we picked random contiguous sequences of notes until the total number $n$ was
	met and we labeled each of the chosen region as ``missing'' or ``extra''
	according to two random variables $p_1$ and $p_2$ defined by a uniform
	distribution in $(0.25, 0.75)$ and $p_2=1-p_1$.

	\begin{figure*}
		\center
		\includegraphics[width=\textwidth]{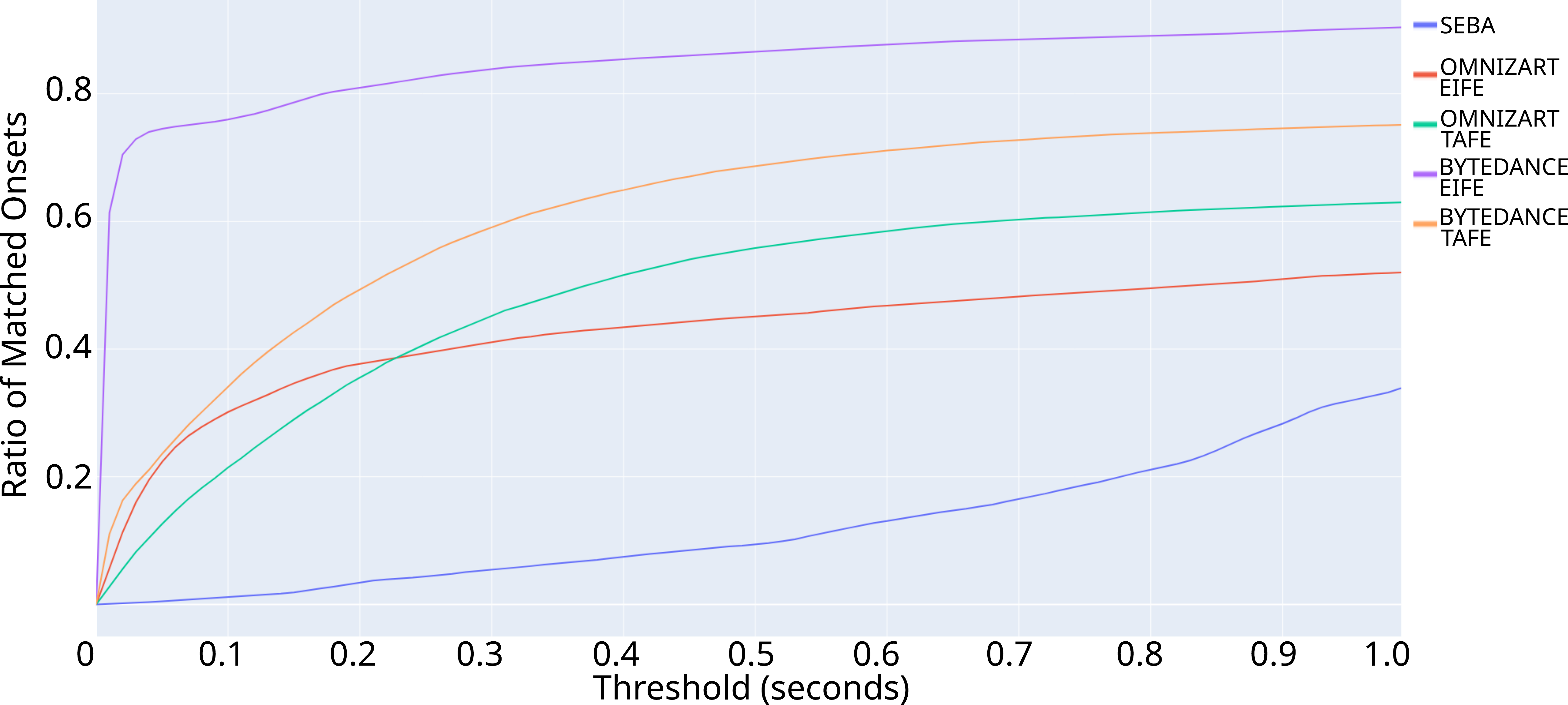}
		\null\vfill
		\includegraphics[width=\textwidth]{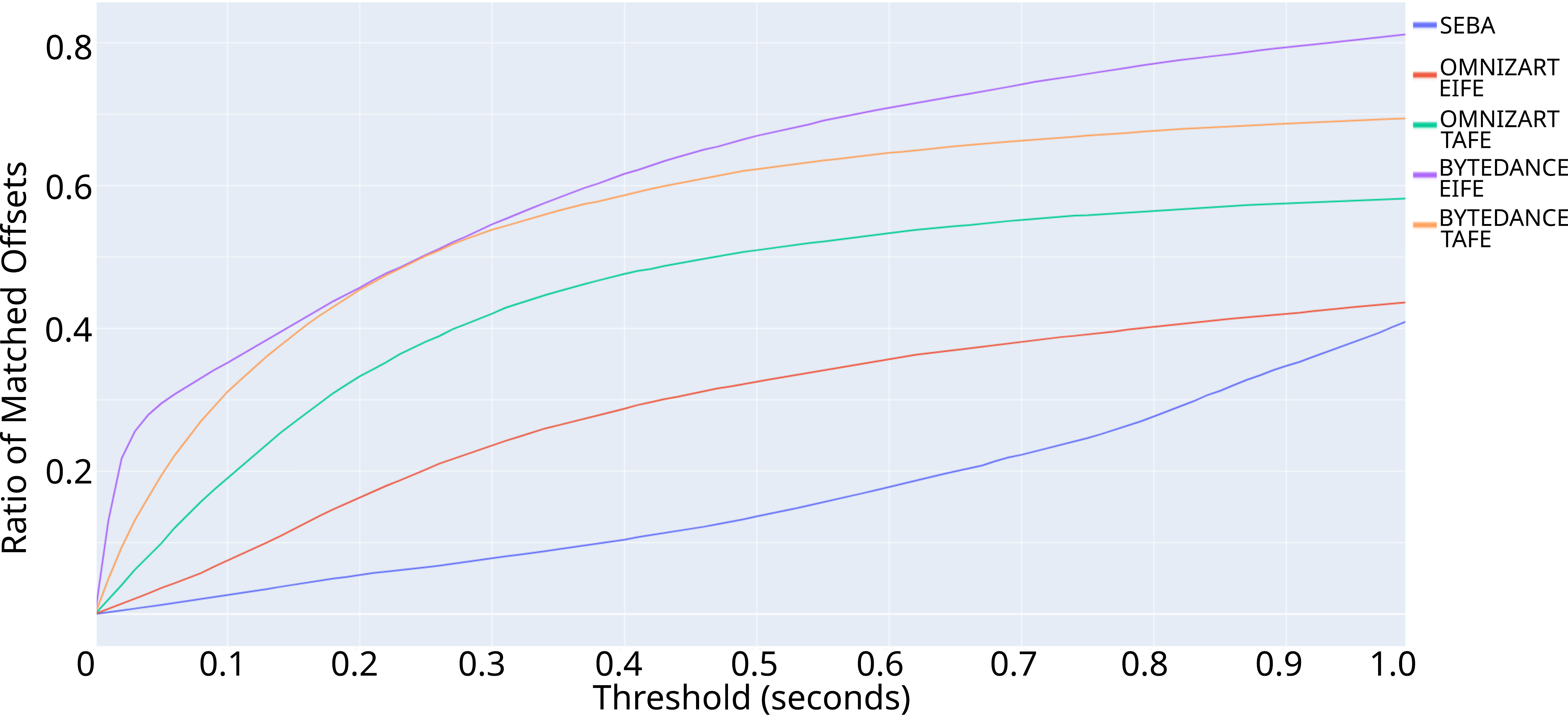}
		\null\vfill
		\caption{
			\added{
				Evaluation on piano-solo music (SMD dataset) without
				missing/extra note. Curves are the ratio macro-averaged curves of ratios
				between the number of matched notes at a given threshold and the total
				number of notes.
			}
		}
		\label{fig:piano_nomissing}
	\end{figure*}

	\begin{figure*}
		\center
		\includegraphics[width=\textwidth]{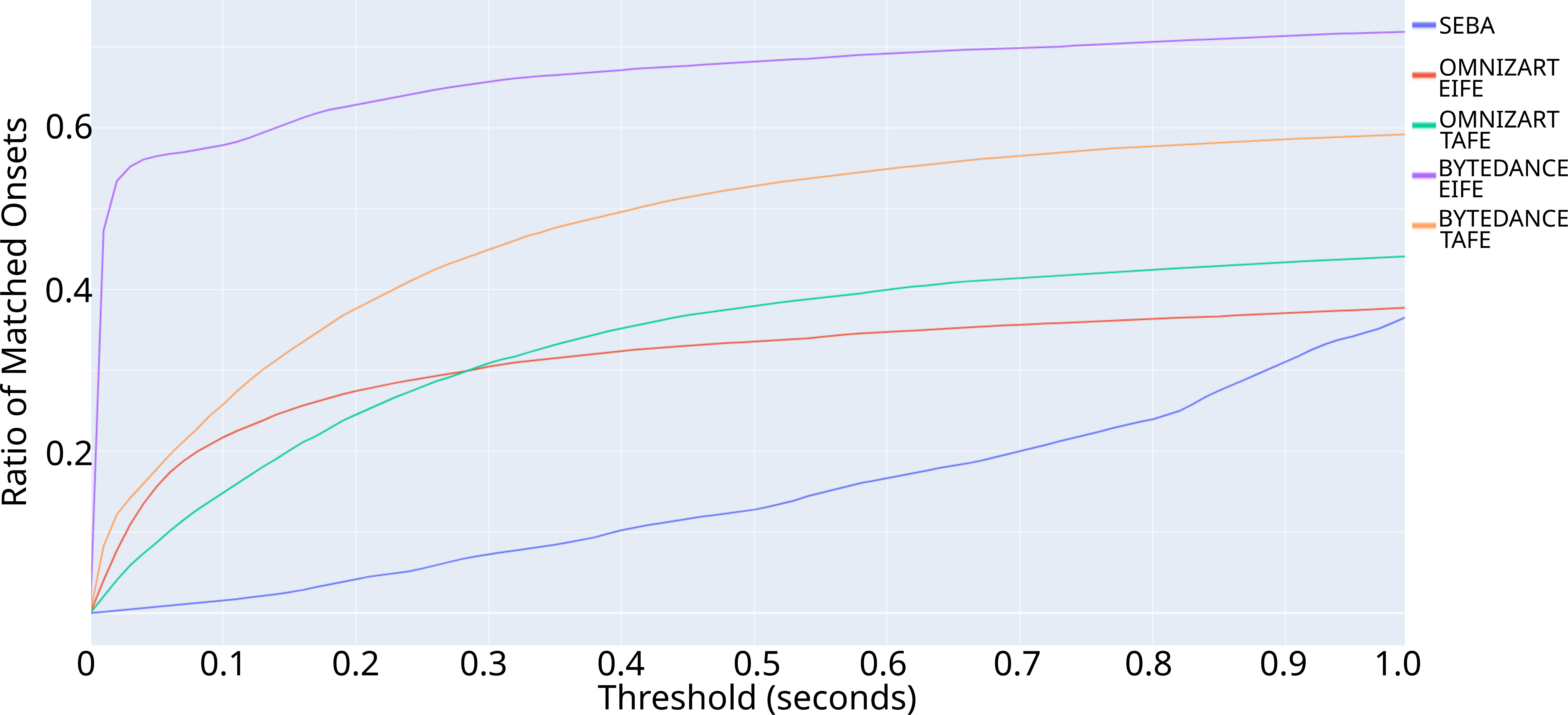}
		\null\vfill
		\includegraphics[width=\textwidth]{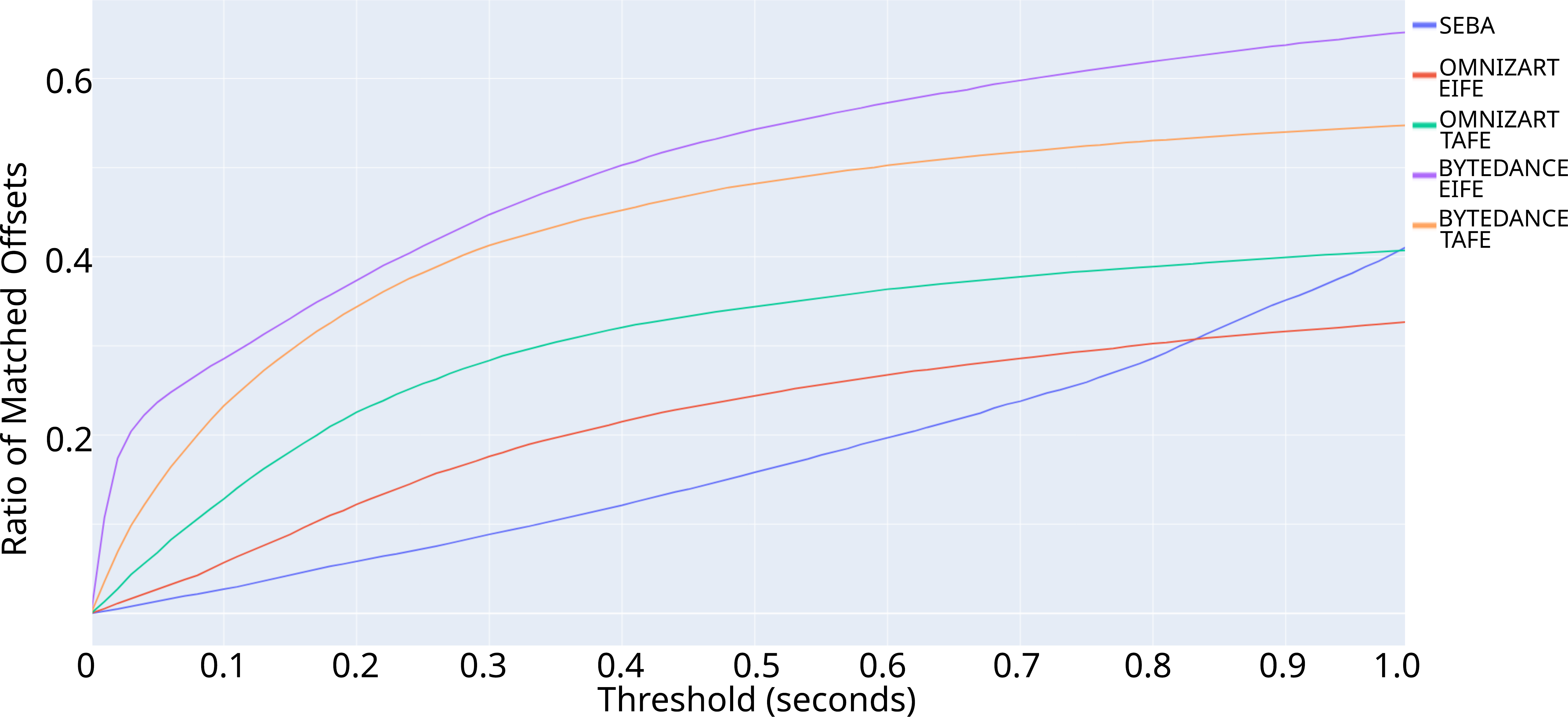}
		\null\vfill
		\caption{
			\added{
				Evaluation on piano-solo music (SMD dataset) with
				missing/extra note. Curves are the ratio macro-averaged curves of ratios
				between the number of matched notes at a given threshold and the total
				number of notes.
			}
		}
		\label{fig:piano_missing}
	\end{figure*}

	\begin{figure*}
		\center
		\includegraphics[width=\textwidth]{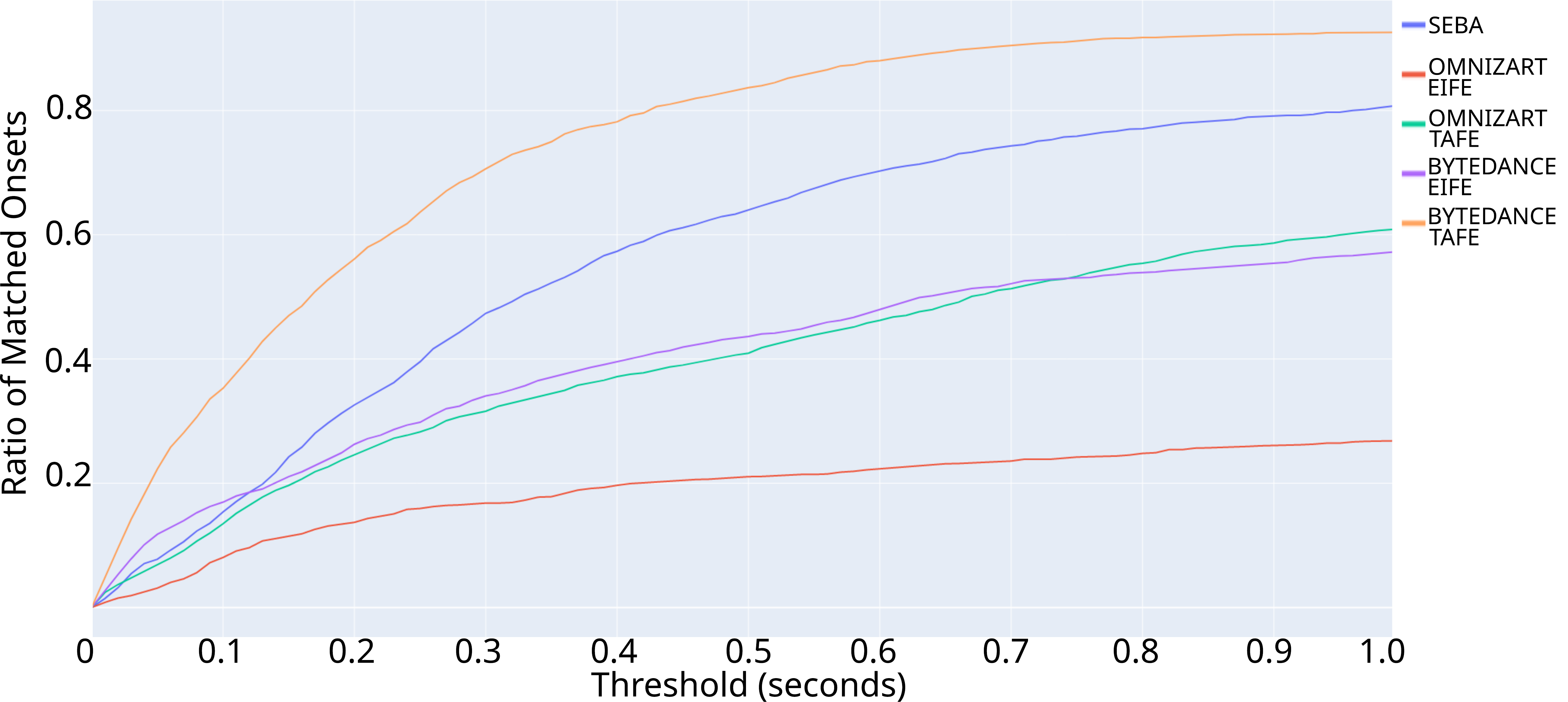}
		\null\vfill
		\includegraphics[width=\textwidth]{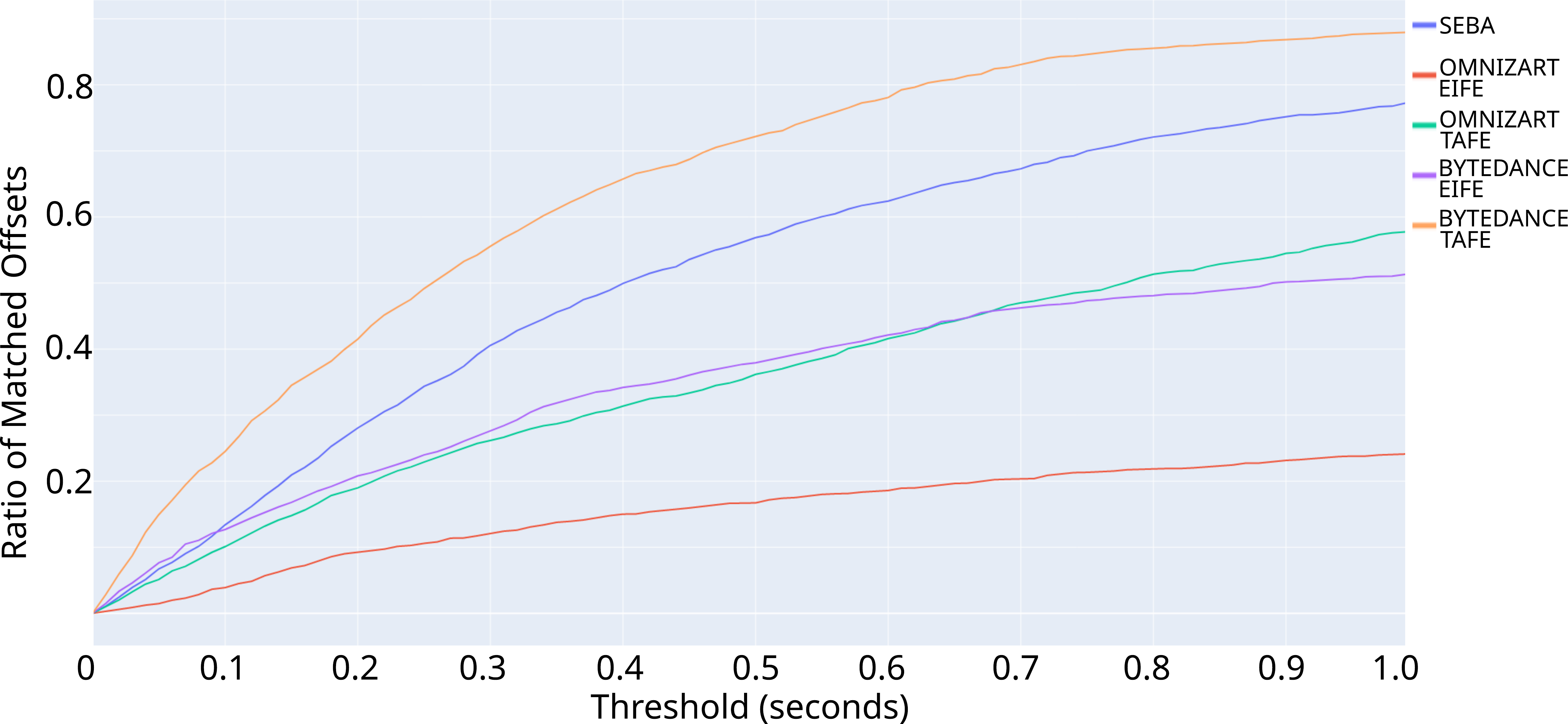}
		\null\vfill
		\caption{
			\added{
				Evaluation on multi-instrument music (Bach10 dataset) without missing/extra
				note. Curves are the ratio macro-averaged curves of ratios between the
				number of matched notes at a given threshold and the total number of notes.
			}
		}
		\label{fig:multi_nomissing}
	\end{figure*}

	\begin{figure*}
		\center
		\includegraphics[width=\textwidth]{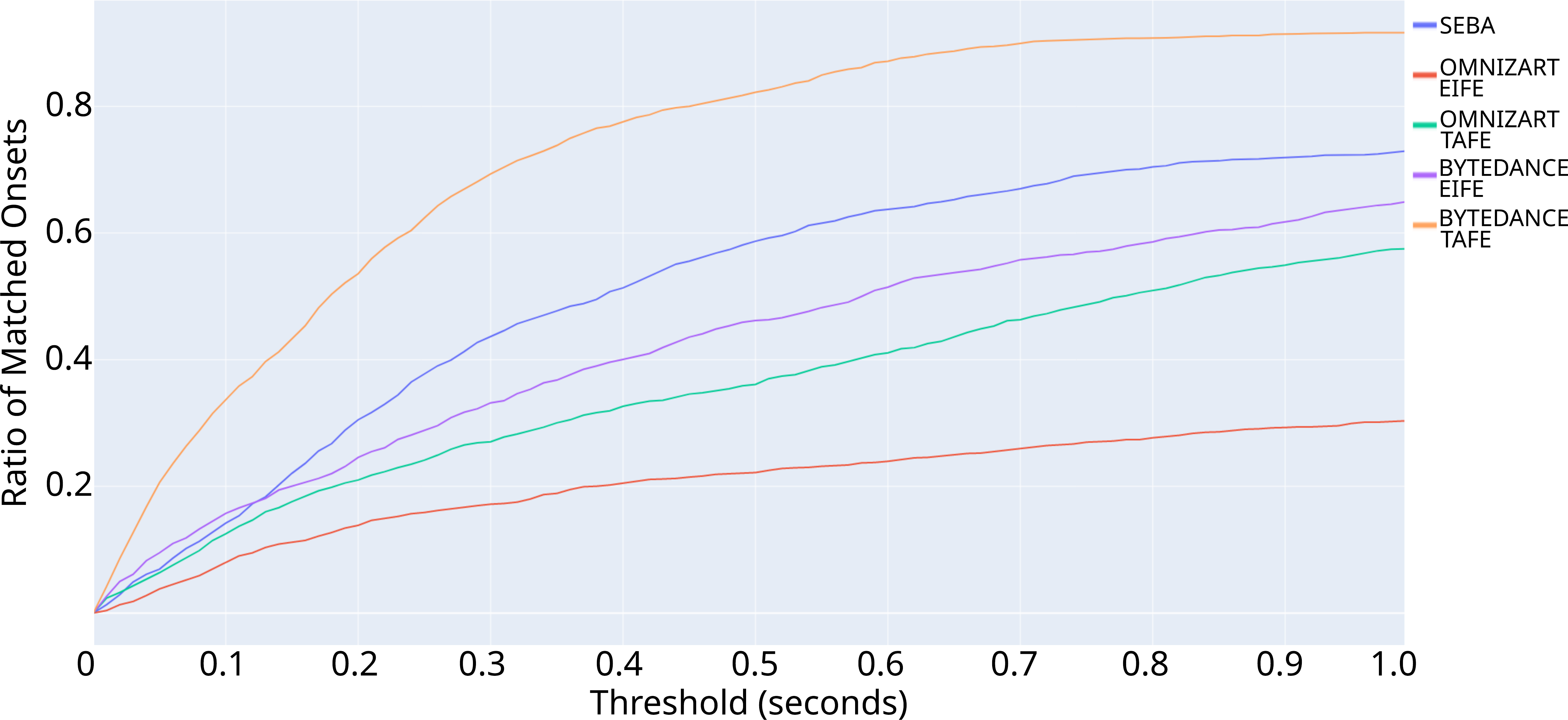}
		\null\vfill
		\includegraphics[width=\textwidth]{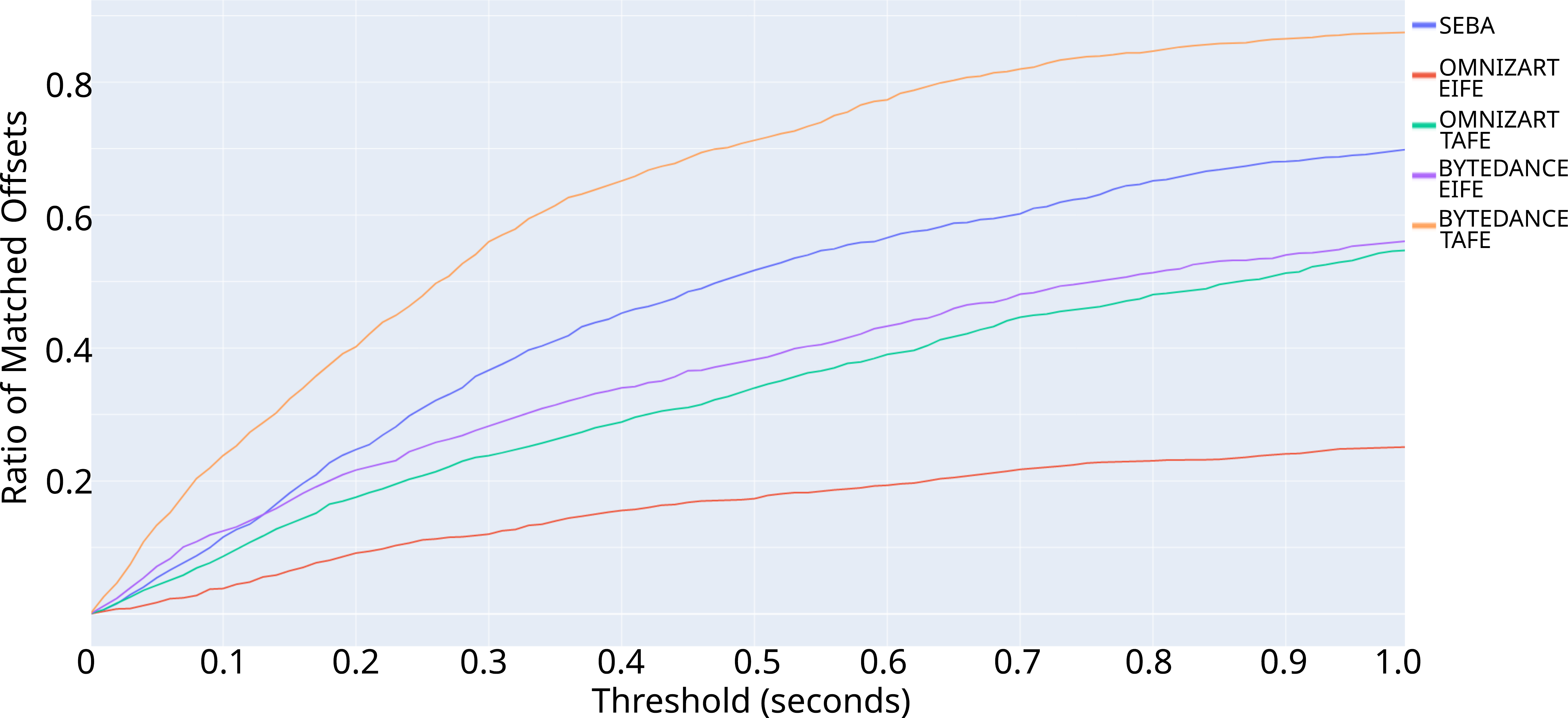}
		\null\vfill
		\caption{
			\added{
				Evaluation on multi-instrument music (Bach10 dataset) with missing/extra
				plots) and with  missing/extra
				note. Curves are the ratio macro-averaged curves of ratios between the
				number of matched notes at a given threshold and the total number of notes.
			}
		}
		\label{fig:multi_missing}
	\end{figure*}

	\section{Experimental Set-Up}\label{sec:experimental_setup}

	We conducted four experiments to cover every aspect of the problem space as
	generated by the combination of two different conditions. Namely, we observed
	how alignment methods change in case
	\begin{enumerate*}[(1)]
		\item missing/extra notes between the score and the performance are
		introduced, and \item instruments other than piano are present.
	\end{enumerate*}
	To ensure a fair comparison of AMT-based A2SA methods, we used two state-of-art
	models, namely one trained on Piano solo music~\cite{kong2020highresolution}
	(BYTEDANCE) and one trained on ensemble music~\cite{wu2020multiinstrument}
	(OMNIZART).

	We used the ASMD Python API to retrieve missing and extra notes computed as
	explained in section \ref{sec:datasets}. To simulate notes unavailable in the
	score, we removed the ``extra'' notes from the artificially misaligned score,
	while to simulate notes not played in the recording --- ``missing'' ---, we
	generated ad-hoc notes using the same procedure and removed them from the
	transcribed performance.  However, since the SEBA method does not rely on AMT,
	it is tested without extra notes. Note that even though we remove notes in the
	input data, we still have them in the ground-truth, allowing to correctly assess
	all inferred timing.

	We also used the ASMD Python API to select the proper datasets for our
	experiments. To avoid over-fitting during the evaluation stage, we did not use
	the Maestro~\cite{hawthorne2019enabling} and
	MusicNet~\cite{thickstun2018invariances} datasets because the AMT models were
	trained on them. Instead, we used the ``Saarland Music
	Dataset''~\cite{muller2011saarland} for evaluating piano A2SA. It consists of 50
	piano audio recordings along with the associated MIDI performances, recorded
	with high-quality piano equipped with MIDI transducers.
	As regards to multi-instrument music, we relied on another well known dataset:
	the ``Bach10''~\cite{duan2011soundprism} dataset, which includes 10 different
	Bach chorales synthesized with virtual chamber instruments. Even though Bach10
	dataset provides non-aligned scores, we used our artificially misaligned data to
	obtain results comparable with the other datasets.

	To reduce the computational cost, we constrained each method inside 32 GB of RAM
	and 600s. Whenever a method failed for an out-of-ram/out-of-time error, the
	specific piece was removed from the evaluation. In doing so, we also get a rough
	reliability estimation of the various approaches here tested. Hence, due to the
	high resources required by EITA, the SMD dataset size is reduced to \added{26}
	music pieces when testing without missing/extra notes and \added{31} music
	pieces when considering them.

	To ease alignment, we preprocessed both score and audio by stretching the note
	timings so that the score duration was the same as the trimmed audio. This
	operation corresponds to enforcing in the music score the performance average
	BPM.

	We tuned the TAFE method by using the 5\% of the available pieces sampled with a
	uniform distribution from the entire ASMD. We ran the TAFE method to find the
	best parameters for aligning the misaligned data to the ground-truth
	performance, after having removed missing and extra notes. We adopted a Bayesian
	Optimization approach with an Extra Trees surrogate model, Expected Improvement
	acquisition function, and exploitation-exploration factor set to 0.01. We used
	180 calls and let the space of parameters being defined by 7 different distance
	functions and the \textit{FastDTW} radius in $[1, 200]$. We found
	\textit{cosine} distance and radius 178 as optimum parameters.  We then
	used the same radius size for \textit{FastDTW} in the EIFE method, while using
	the distance defined by SEBA.

	For synthesizing MIDI files in the SEBA and EIFE methods, we employed the
	freely available MuseScore SoundFont\footnote{MuseScore 2.2 SF2 version:
		\url{https://musescore.org/en/handbook/3/soundfonts-and-sfz-files}}. As
	evaluation measure, we observed the ratio of matched onsets and offsets under
	several different thresholds in each music piece; we then averaged the obtained
	curves  to get a macro-curve representing the overall performance of each
	method.

	\begin{figure*}
		\center
		\includegraphics[width=0.65\textwidth]{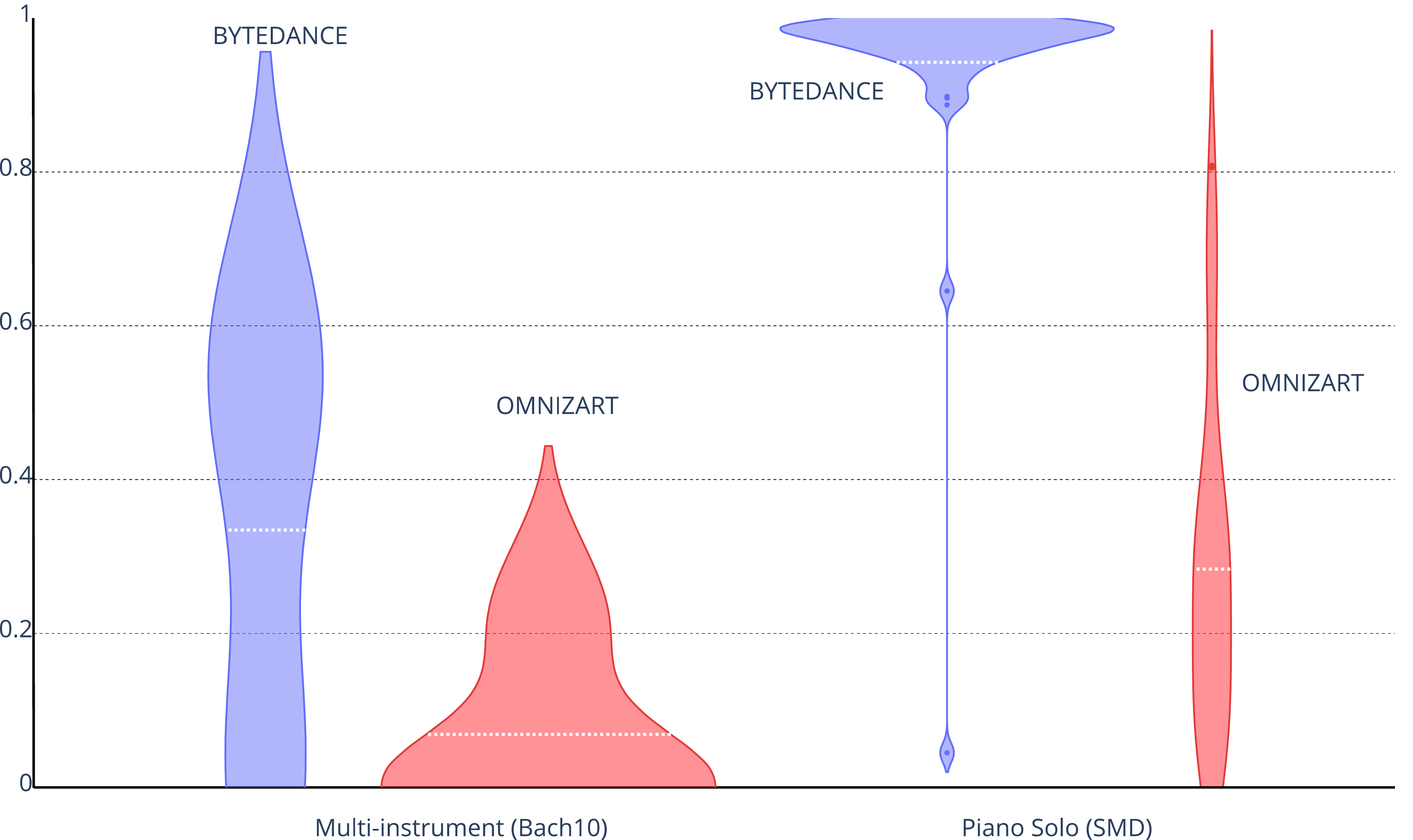}
		\caption{Onsets F1-measure for the two AMT models}
		\label{fig:f1measure}
	\end{figure*}

	\section{Experimental Results}
	\label{sec:results}

	With regard to piano-solo music, methods based on BYTEDANCE model are
	outperforming the rest -- see
	Fig.~\ref{fig:piano_nomissing}~and~\ref{fig:piano_missing}. EIFE manages to
	exploit the good onset prediction of BYTEDANCE better than TAFE. However, the
	performance decreases when considering offsets due to the poor inference of
	onset positions of generic AMT models.  OMNIZART does not perform well, as shown
	in Figure~\ref{fig:f1measure}, which is expected as it is trained on
	multi-instrument music. Finally, SEBA method seems more robust in offset
	prediction and maintains a similar score for both offsets and onsets.

	\added{
		Furthermore, we observe that in non-piano music TAFE method is the
		best-performing one -- see
		Fig.~\ref{fig:multi_nomissing}~and~\ref{fig:multi_missing}. Indeed, the good
		performance of AMT models, makes EIFE and TAFE approaches still reliable,
		especially for little thresholds -- i.e. $< 0.1$ seconds. Moreover, even though
		we were expecting a useful input from OMNIZART multi-instrument model, we
		observed better performance with BYTEDANCE; this could be due to the low
		generalization ability of OMNIZART --- see Fig.~\ref{fig:f1measure}.
	}

	Every considered model suffers in case of missing notes, while retaining a
	similar curve shape and proportions. As such, we think that the most promising
	option for increasing the performance of A2SA with missing and extra notes is to
	increase the overall alignment accuracy.

	\section{Conclusion}
	\label{sec:conclusions}

	We designed a methodology to compare various alignment systems and proposed two
	methods for frame and note-level alignment. After extensive experiments, it was
	shown that the proposed method for note-level alignment brings notable
	advancement to the state-of-art thanks to the AMT models. \added{Moreover, even
		if AMT is still not reliable for non-piano solo music, the top-performing
		approach among those tested is still based on the AMT model trained on
		piano-solo music. Our intuition is that the size of the training dataset is
		extremely relevant for the good performance of the model.}

	Our future studies will focus on the quality of the alignment with perceptual
	measures to confirm the results obtained through the present assessment.

	\balance
	\bibliographystyle{jabbrv_ieeetr}
	\bibliography{./IEEEabrv.bib,bibliography.bib}
\end{multicols}

\end{document}